\documentstyle[epsf]{mn}
\newcommand{\av}{$A_{V}$}

\newcommand{\ebv}{ {\it E(B--V)}}

\newcommand{\bb}{\bibitem[]{bla}}
\newcommand{\zm}{ \relax \ifmmode {\rm M_{\odot}} \else {M$_{\odot}$}\fi}

\newcommand{\ang}{$\rm \AA$}

\newcommand{\lm}{L$_{\odot}$}
\newcommand{\rsun}{R$_{\odot}$}

\newcommand{\degree}{$^{\rm o}$}
\newcommand{\arc}{$^{\prime\prime}$}
\newcommand{\arcm}{$^{\prime}$}
\newcommand{\mic}{$\mu$m}
\newcommand{\ea}{{et al.}}

\newcommand{\km}{km s$^{-1}$}
\newcommand{\ha}{H$\alpha$}
\newcommand{\hb}{H$\beta$}

\def\mv{\mbox{$m_{_V}$}}

\def\lesssim{\mathrel{\hbox{\rlap{\hbox{\lower4pt\hbox{$\sim$}}}\hbox{$<$}}}}
\def\gtrsim{\mathrel{\hbox{\rlap{\hbox{\lower4pt\hbox{$\sim$}}}\hbox{$>$}}}}

\def\arcmin{\hbox{$^\prime$}}
\def\arcsec{\hbox{$^{\prime\prime}$}}

\def\ion#1#2{#1$\;${\small\rm\@Roman{#2}}\relax}
 
\newcommand{\lam}{$\lambda$}
\newcommand{\fb}[2]{[#1\,{ \sc #2}]}

\newcommand{\al}[2]{#1\,{\sc #2}}
\newcommand{\Av}{$A_{V}$}

\newcommand{\hea}{HD~87643}
\newcommand{\Rv}{R$_V$}

\newbox\grsign \setbox\grsign=\hbox{$>$} \newdimen\grdimen 
\grdimen=\ht\grsign
\newbox\simlessbox \newbox\simgreatbox
\setbox\simgreatbox=\hbox{\raise.5ex\hbox{$>$}\llap
     {\lower.5ex\hbox{$\sim$}}}\ht1=\grdimen\dp1=0pt
\setbox\simlessbox=\hbox{\raise.5ex\hbox{$<$}\llap
     {\lower.5ex\hbox{$\sim$}}}\ht2=\grdimen\dp2=0pt
\def\simgreat{\mathrel{\copy\simgreatbox}}
\def\simless{\mathrel{\copy\simlessbox}}

\makeatletter 
\renewcommand\@biblabel[1]{}     

\makeatother

\begin{document}

\title[HD 87643]
{The evolved B[e]  star  HD 87643 :
observations and a radiation driven disk-wind model for B[e] stars
}

\author[Ren\'e D. Oudmaijer \ea ]
{ Ren\'e D. Oudmaijer$^{1}$, Daniel Proga$^{1}$, Janet E. Drew$^{1}$, Dolf de Winter$^{2,3}$ \\
$^{1}$ Imperial College of Science, Technology and Medicine,
Blackett Laboratory, Prince Consort Road,\\ London,  SW7 2BZ, U.K.   \\
$^{2}$ 
Dpto. F\'{\i}sica T\'eorica C--XI, Facultad de Ciencias, Universidad
Aut\'onoma de Madrid, Cantoblanco, E--28049 Madrid, Spain \\
$^{3}$ Centro de Astrof\'{\i}sica de Universidade do Porto, Campo Alegre 823, 
4150 Porto, Portugal \\
}

\date{received,  accepted}

\maketitle
\begin{abstract}

New high resolution spectroscopic and medium resolution
spectropolarimetric data, complemented with optical broad and narrow
band imaging, of the B[e] star HD 87643 are presented.  The spectrum
of HD 87643 exhibits the hybrid characteristics well known to be
representative of the group of B[e] stars; a fast wind with an
expansion velocity in excess of 1000 \km\ is measured in the hydrogen
and helium lines, while a slower component is traced by lower
excitation lines and forbidden lines.  Clues to the geometry of the
rapidly expanding circumstellar shell are provided by the startling
polarization changes across \ha.  Comparison with published schematic
calculations indicates that the polarizing material is located in a
slowly rotating, expanding disk structure.  A hydrodynamical model is
then presented whose results are consistent with the original two-wind
concept for B[e] stars and exhibits kinematic properties that may well
explain the observed spectral features in HD 87643.  The model
calculations use as input a B star undergoing mass loss, surrounded by
an optically thick disk.  The resulting configuration consists of a
fast polar wind from the star and a slowly expanding disk wind.  The
model also predicts that the stellar wind at intermediate latitudes is
slower and denser than in the polar region.

\end{abstract}


\section{Introduction}

The B[e] phase is one of the phases a massive star might go through
during its evolution. Since the spectra of such objects are dominated
by emission lines, and generally do not show any signs of photospheric
absorption lines, it was only due to the work of Zickgraf \ea\ (1985,
1986, 1996 and references in these papers) on peculiar B-type emission
line stars in the Magellanic Clouds that the luminosities of some of
these objects began to be known, and their location in the HR-diagram
could be plotted with some accuracy.

\hea\ ($\alpha_{2000}$ = 10$^h$04$^m$30$^s$, $\delta_{2000}$ =
--58\degree38\arcm52\arc ; $l$=283\degree, $b$=--2.5\degree) appears
to be a Galactic counter-part of the B[e] stars in the Magellanic
Clouds.  Over the last decades spectroscopic studies of the object
were reported by Hiltner, Stephenson \& Sanduleak (1968), Stephenson
(1974), Swings (1974) and Carlson \& Henize (1979).  The picture that
has emerged is that the optical spectrum of HD 87643 is dominated by
emission lines from Fe {\sc ii} and the Balmer series, and displays
low excitation forbidden lines.  Except for blueshifted P Cygni
absorption components and interstellar lines no optical absorption
lines have been detected.  Optical and ultraviolet IUE spectra of
\hea\ are presented by de Freitas Pacheco, Gilra \& Pottasch (1982)
and de Freitas Pacheco \ea\ (1985).  The UV spectrum appears to be
dominated by low ionization absorption lines, although features due to
Fe {\sc iii}, Si {\sc iii} and Si {\sc iv} are present.

McGregor, Hyland \& Hillier (1988) find that \hea\ displays a large
near-infrared excess ({\it J} -- {\it K} $\sim$ 2.6), which they
attribute to hot circumstellar dust.  Published broad band photometry
indicates variations over long timescales: the {\it V} band magnitude
was reported as 8.51 in 1968 (Hiltner \ea\ 1968), and as 9.41 in 1995
(Torres \ea\ 1995).  In the intervening years, the {\it V} band
magnitude has fluctuated, but an overall trend towards fainter
magnitudes is present (Miroshnichenko 1998). No colour information has
been published however.  The emission line spectrum of the star is
also variable.  Lopes, Daminelli Neto \& de Freitas Pacheco (1992)
show the behaviour of the \hb\ line.  Its equivalent width (EW)
changed by a factor of several in the period 1980 -- 1990, reaching a
maximum value in 1985.

An interesting aspect of the surroundings of \hea\ is discussed by
Surdej \ea\ (1981) and Surdej \& Swings (1983).  They obtained spectra
of two high density knots in the large reflection nebula around \hea\ at
offsets from the star of 20\arc\ NW and 17\arc S.  Surdej \& Swings
(1983) show that the central peak \hb\ emission at the two positions
differs by about 150 \km, and argue that the observed velocity
difference confirms their hypothesis that the line forming region is not
isotropic. 

The layout of this paper is as follows; Sec.~2 describes our
observations of the star.  Sec.~3 describes key features of the
optical spectrum, and then Sec.~4 discusses the polarization of the
object. Constraints on the distance and hence on the evolutionary
status of the object are provided in Sec.~5.  Then in Sec.~6 a new
dynamical model for the inner circumstellar environment of HD 87643
and B[e] stars in general is outlined.  A virtue of this model is that
it offers great promise of fitting in with the striking
spectropolarimetric behaviour at \ha\ seen in \hea .  The paper's
conclusions appear in Sec.~7.

\section{Observations}

\subsection{Echelle spectroscopy}

Spectroscopic data were obtained during the nights of 20 and 21 April
1997 employing the Coud\'e echelle spectrograph on the 74\arcsec\
telescope of  Mount Stromlo Observatory. 
The observational set-up included a 2 arcsec slit, the 300 lines/mm
cross-disperser grating and a 2K $\times$ 2K TEK CCD. Since the
cross-disperser is not located in the pupil plane, vignetting occurs
away from the central order. Hence, the CCD was windowed during
read-out. To improve signal-to-noise, the data were binned over two
pixels in both the dispersion and cross-dispersion directions.  

On 20 April, observations at a central wavelength of 6562 \ang\ were
obtained.  The total on-source integration time was 45 minutes, obtained
in 3 individual exposures of 15 minutes each.  Observations of a bright
B star (HR 4199, B0V) were also obtained to correct the spectrum for
telluric absorption lines, and Th-Ar arc maps were obtained for the
dispersion correction.  The dispersion fit through more than 450
identified lines yielded an rms value of 0.009 \ang\ using 4$^{\rm th}$
order polynomials in both the dispersion and cross-dispersion direction. 
The final spectrum consisted of 20 different orders covering the range
from 5750 to 7220 \ang, with a spectral resolution as measured from
telluric absorption lines of 8.5 \km.  The data were corrected for
telluric absorption by matching the strength of the absorption lines
with those in HR 4199. 

On 21 April, two 15 minute exposures in the blue at a central
wavelength 5000 \ang\ were obtained.  The spectra contained 33 orders,
and covered a wavelength range from 4270 -- 5705 \ang . Due to shorter
exposure times and the reddening of the object, the blue spectrum has
a lower signal-to-noise (SNR) than the red spectrum.

Data reduction was performed in {\sc iraf} (Tody, 1993), and included
the steps of bias subtraction, flatfielding, background subtraction and
wavelength calibration.

\subsection{Spectropolarimetry}

Optical linear spectropolarimetric data were obtained using the RGO
Spectrograph on the 3.9-metre Anglo-Australian Telescope on 31
December 1996 and 1 January 1997.  The weather was clear.

The instrumental set-up included a half-wave plate rotator set to
various angles to obtain the Stokes QU parameters and a calcite block to
separate the light leaving the retarder into perpendicularly polarized
light waves.  Two holes in the dekker allow for simultaneous observations
of the object and the sky.  Four spectra are recorded, the O and E rays
of the target object and the sky respectively.  One complete
polarization observation consists of a series of consecutive
exposures at four rotator positions.  Spectropolarimetric standards
were observed during both nights.  The instrumental polarization was
found to be negligible. 

A 1024 $\times$ 1024 pixel TEK-CCD detector was used which, combined
with the 1200V grating, yielded a spectral coverage from 6340 -- 6880
\ang, in steps of 0.54 \ang.  Wavelength calibration was performed by
observing a Cu-Ar lamp before each exposure.  On 31 December 5
$\times$ 4 sets of data were taken with exposure times of 90~s, a day
later 3 $\times$ 4 sets with exposure times of 120~s were obtained.  A
slit width of 1.5 arcsec was used.

The initial processing of the data was performed in the {\sc iraf}
package.  After bias-subtraction and flatfielding, the individual
spectra were extracted by tracing the pixel rows containing the data. 
The wavelength calibration was performed by fitting a 3$^{\rm rd}$ order
polynomial through more than 20 identified lines in the arc spectra, and
resulted in dispersion fits with a $\sigma$ of 0.07 \ang.  The resulting
spectral resolution as measured from arc lines is 60 \km.  The E
and O ray data were then extracted and imported
into the Time Series/Polarimetry Package ({\sc tsp}) incorporated in the
{\sc figaro} software package maintained by {\sc starlink}.  The Stokes
parameters were determined and subsequently extracted.

\subsection{Imaging}

Images of HD 87643 were obtained through {\it V}, {\it R}, \ha\ and
[S\,{\sc ii}] filters with the Dutch 92~cm telescope stationed at the
ESO La Silla.  On 12 July 1992 images were taken through the {\it R}
(ESO\#421), H${\alpha}$ (ESO\#387) and [S\,{\sc ii}] (ESO\#391)
pass-bands for 4\,${\times}$\,3, 3\,${\times}$\,30 and
3\,${\times}$\,70 sec, respectively.  The detector used was a GEC
576\,${\times}$\,385-pixel CCD. The pixels have a size of 22 \mic\
corresponding to 0.36 arcsec on the sky.  On 20 April 1995 six
images of 10~sec were obtained with the ESO\#420 {\it V} filter.  In
this case, the CCD was a Tektronix 512\,${\times}$\,512 pixels CCD,
with a pixelsize of 27 \mic\ corresponding to 0.44 arcsec.

After the images were corrected for bias, cleaned of cosmic ray hits, and
flatfield-corrected, the images were aligned and co-added.  The seeing 
measured from Gaussian fits to field-star images corresponded to a 
full-width-at-half-maximum (FWHM) of 1.8 arcsec.

\section{The optical spectrum}

\begin{figure}
\mbox{\epsfxsize=0.48\textwidth\epsfbox{
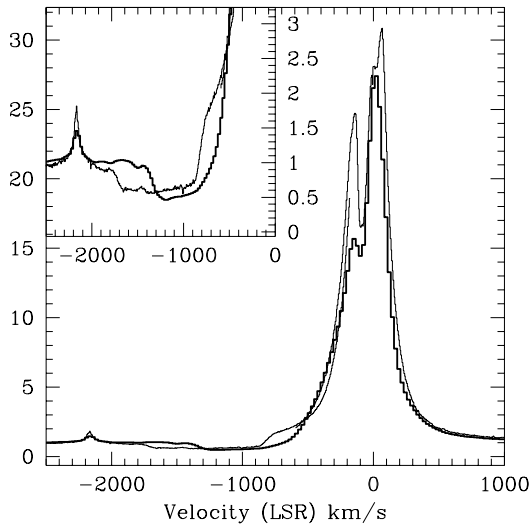
}}
\caption{
The \ha\ line  in January (thick line) and April 1997 (thin line). 
The insert in the top left-hand corner is a magnification around the P Cyg 
absorption. The bluest emission peak is the \al{Fe}{ii} \lam 6517 line. 
\label{ha}
}
\end{figure}

\begin{figure}
\mbox{\epsfxsize=0.48\textwidth\epsfbox{
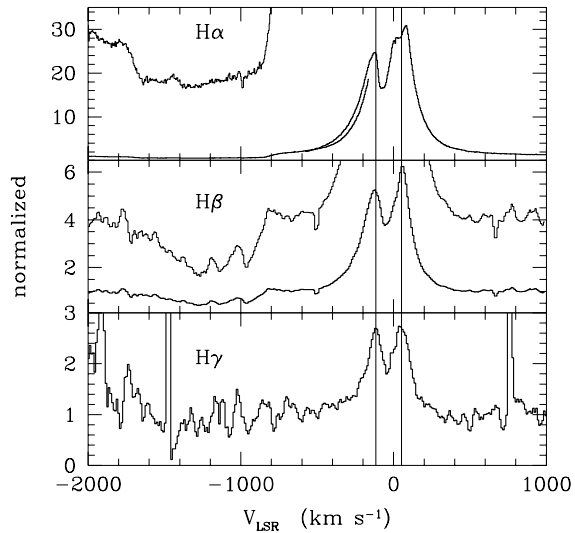
}}
\caption{
The hydrogen recombination lines. For clarity, the \ha\ and H$\beta$
lines have been multiplied by 30 respectively 4 to enhance the 
absorption. The vertical lines indicate the central velocities of the
\ha\ emission peaks.
\label{hyd}
}
\end{figure}

\begin{figure}
\mbox{\epsfxsize=0.48\textwidth\epsfbox{
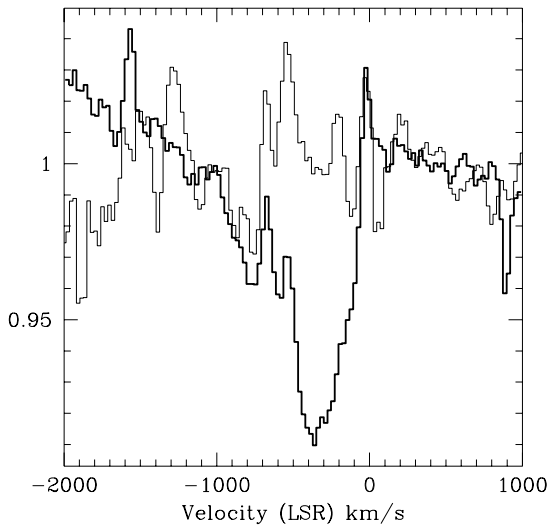
}}
\caption{
As Figure~\ref{ha}, but now for the \al{He}{i} \lam 6678 line.  The
April spectrum has been smoothed to the same resolution as the January
spectrum.  The different slopes at the blue end of the spectra are the
result of the continuum flattening; the line is located in the red
emission wing of \ha.  The emission feature at $\sim$ --1600 \km\ in
the January data is real, as it appears in both the 31 December 1996
and 1 January 1997 spectra, the line could however not be identified.
\label{he}
}
\end{figure}

\begin{figure*}
\mbox{\epsfxsize=0.9\textwidth\epsfbox[40 180 530 430]{
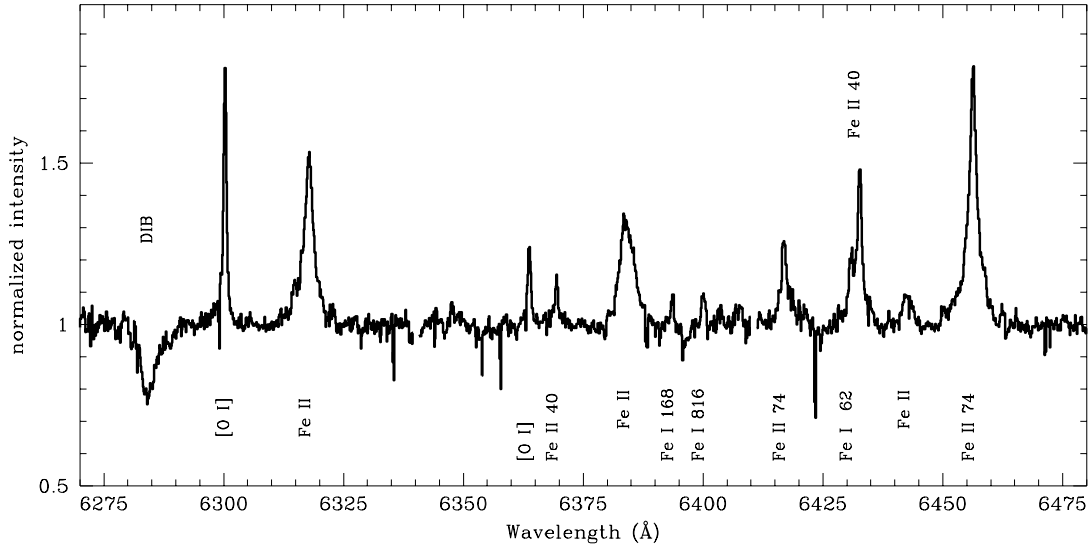
}}
\caption{
A representative part of the April 1997 spectrum, rebinned to 0.15 \ang,
containing the many different line profiles in the spectrum of HD 87643.
\label{ofe}
}
\end{figure*}

The optical spectrum is dominated by permitted and forbidden emission
lines of Fe {\sc ii}, and the Balmer lines.  Some other lines are
visible, most notably lines of [O {\sc i}] and several strong \al{Fe}{i}
lines.  In some He {\sc i} lines emission is accompanied by broad
absorption.  Diffuse interstellar bands (DIBs) are present in
absorption.  In Figs.~\ref{ha}, \ref{hyd}, \ref{he} and \ref{ofe},
several parts of the spectrum are shown.  What catches the eye are the
very broad blueshifted absorption components in \ha, \hb\ and the
He {\sc i}~$\lambda$ 6678 line.  The \fb{O}{i} lines are markedly
narrower than the \al{Fe}{ii} lines.  We find a Local Standard of Rest
(LSR) velocity of --17 $\pm$ 4 \km\ from the Fe {\sc ii} and \fb{O}{i}
lines.  We take this figure to be indicative of the system's radial
velocity. 

The emission in \ha\ is double-peaked and displays broad wings.  The
broad P Cygni type absorption extends to an unprecedented blueshifted
edge velocity of 1800 \km\ with respect to the systemic velocity.  The
broad absorption in the He {\sc i} $\lambda$6678 line indicates that
this high excitation line is also formed in a circumstellar outflow.
As there are no bona fide photospheric absorption lines present in the
spectrum, it is possible that there is a non-stellar contribution to the
optical spectrum.  The dearth of photospheric absorption lines extends
into the ultraviolet.  Based on the many low excitation lines, de
Freitas Pacheco \ea\ (1985) suggested that the UV spectrum is due to a
`cool' wind.

\subsection{The Balmer lines}

Fig.~\ref{ha} shows an overplot between the January 1997 (thick line)
and the April 1997 (thin line) observations. As the free spectral
range of the echelle spectrograph was smaller than the total line
width of \ha, we show an overplot of the two orders containing the
\ha\ line. Despite problems with the continuum rectification, which
was performed independently of the orders covering \ha, the agreement
in the overlap region is very good.

The strength of especially the blue emission peak in \ha\ has
appreciably increased in April 1997, as has the strength of the Fe{ \sc
ii} line at 6517 \ang.  Perhaps more interesting are the changes in
the edge velocities of the \ha\ absorption component.  Whereas in
January 1997, the epoch of our spectropolarimetric observations, the
maximum absorption blueshift was of order 1500 \km, reaching in to a
minimum velocity of $-$600\km, this has changed to a velocity range of
$-$1800 to $\sim -$800 \km\ in April.  The changes are undoubtedly
real.  Also, since the high resolution April data oversample the
spectrum, the edges to the P~Cygni absorption are indeed as sharp as
they appear to be.

Fig.~\ref{hyd} shows \ha, H$\beta$ and H$\gamma$ on the same velocity
scales.  The higher order Balmer line spectra have been re-binned to
pixels of size 0.15 \ang\ to increase the SNR of the spectra.  The
P~Cygni absorption is less deep in \hb\ and not necessarily present at
all in H$\gamma$.  The double emission peaks occur at the same
velocities in all three Balmer lines, the velocity difference being 180
\km\ (LSR).  This argues against the common hypothesis that the
double-peaked appearance of the Balmer line emission betrays formation
in a Keplerian circumstellar disk.  One would expect that if rotation
were the dominant kinematic component in the line-forming region, the
velocity difference between the two peaks would increase up the Balmer
series as the line forming region is smaller due to the lower line
opacity.  This is not seen. 

In addition, the central reversal in \ha\ is a little more blueshifted
than that in \hb\ and H$\gamma$, (respectively --60 \km\ (for \ha ) and
--30 \km\ with respect to systemic).  This is also inconsistent with
what would be expected of a purely rotational velocity field.  Instead,
it is possible that the central reversals are due to self-absorption in
a slowly expanding wind.

\subsection{Other spectral lines}

There is a hint of \al{He}{i} \lam 5876 absorption in the April
echelle spectrum, but the low SNR prevents any further comment.
However, the \al{He}{i} \lam 6678 line is clearly present in both
spectra (Fig.~\ref{he}).  The April spectrum has been box-car smoothed
over 15 pixels, and re-binned to pixels of 0.5 \ang\ to enable
comparison with the higher SNR spectrum from January.  In January, the
\lam 6678 line showed a faint narrow emission (at --16 \km)
accompanied by a broad P Cygni type absorption with a terminal
velocity of $\sim$ 1100 \km.  This velocity is somewhat lower than
found in the \ha\ line.  The changes between January and April are
qualitatively like those observed in the \ha\ line; the low velocity
absorption has disappeared in the April data, but there are hints that
both the emission and a part of the high velocity absorption at about
--900 \km\ are still present.

To illustrate the different heavy element emission line profiles that 
are present in the optical spectrum of HD 87643, we show the wavelength
range 6275 -- 6475 \ang\ in Fig.~\ref{ofe}.  Beyond the conspicuous DIB in 
absorption at 6284 \ang , the \fb{O}{i} doublet at 6300 and 6363 \ang, and a 
host of \al{Fe}{ii} lines and some \al{Fe}{i} lines appear.  Both \fb{O}{i} 
lines peak at --14 $\pm$ \ 5\km, with a FWHM of 40 \km, while most \al{Fe}{ii} 
lines have peak velocities between --14 and --22 \km, and are much broader, 
with FWHM of order 100 -- 150 \km.  However these larger FWHM values are under 
half those of the Balmer lines.  Some weak \al{Fe}{ii} lines and the 
\al{Fe}{i} lines have FWHM comparable to the sharp [O{ \sc i}] lines.

Based on the observed line profiles in its spectrum, it would seem
that the circumstellar medium of \hea\ consists of three kinematic
components: firstly, a very strong, high velocity wind with outflow
velocities in excess of 1000 \km\ that the hydrogen and helium lines
sample; secondly, a slower outflow traced by the broad \al{Fe}{ii}
lines and H{\sc i} core emission; and lastly, the narrow forbidden
lines, narrow \al{Fe}{i} lines and some weaker \al{Fe}{ii} lines,
constitute a third component that may be due to a much larger
low-density nebula around the star.

\begin{figure}
\includegraphics{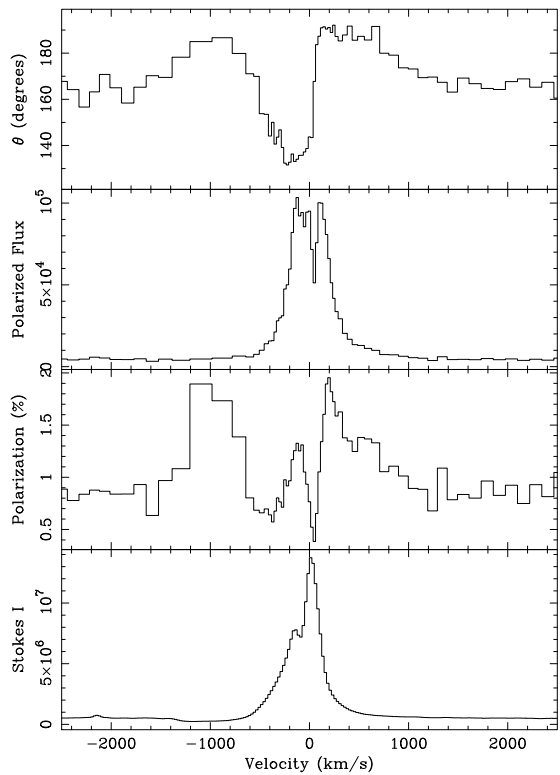}
\vspace{11cm}
\caption{
The polarization spectrum around \ha.  The spectrum has been rebinned 
such that the individual pixels have 1$\sigma$ uncertainties in
polarization of 0.1\% .
\label{pol}
}
\end{figure}

\section{Polarization}

Within the error-bars, the polarization characteristics of the 31
December 1996 and 1 January 1997 spectra are equal. 
The linear polarization spectrum around \ha\ is shown in
Fig.~\ref{pol}.  The panels show, from bottom to top, the total
intensity spectrum, the polarization percentage, the polarized flux
(i.e. intensity times polarization) and the polarization angle.

The observed polarization is enhanced in the P Cygni absorption and
along much of the emission profile.  In the polarized light spectrum,
the enhanced polarization in the absorption component has disappeared. 
Since polarization changes across \ha\ are present, one can already
conclude two things; firstly, the main mechanism that causes the
polarization does not act equally on line and continuum and so must act
within the line forming region, ruling out scattering by small dust
particles, visible in the near-infrared excess.  Secondly, the
projection of this region onto the plane of the sky is not circularly
symmetric. 

Obviously, the observed polarization includes contributions of unknown
magnitude from interstellar polarization and, possibly, circumstellar
dust.  The effects of both on the resulting polarization spectrum are
not simple.  However, the situation presented in QU space (rather than
in the wavelength domain) is easier to grasp since the interstellar
polarization component on its own must add the same (Q,U) vector to all
points.  Circumstellar dust polarization can also be considered constant
over the narrow wavelength range of the spectrum, and so the same
consideration applies. 

\begin{figure*}
\mbox{\epsfxsize=0.9\textwidth\epsfbox{
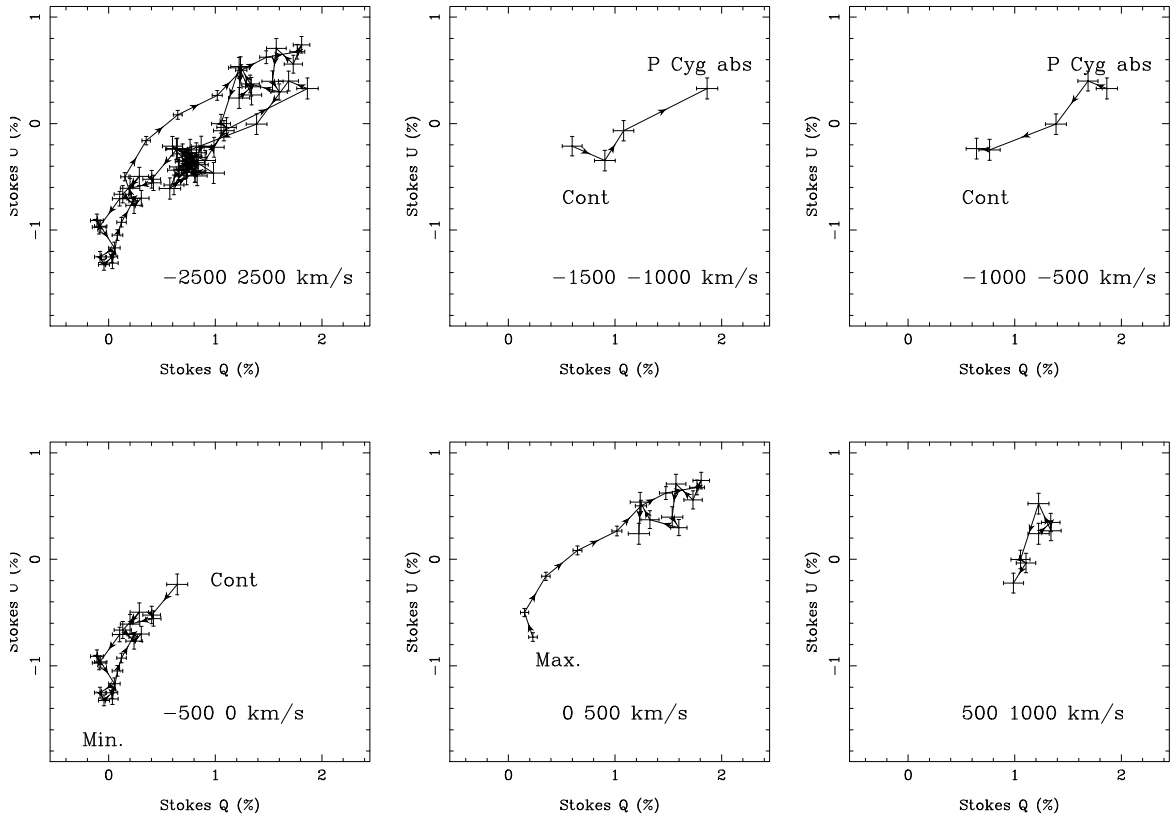
}}
\caption{
The Stokes QU vectors  across \ha\ in HD 87643. The spectra have been
rebinned such that the QU vectors have 1$\sigma$ errors of 0.1\%.
The panels show different parts of the spectrum as function of velocity
(LSR), which are indicated in the bottom of each panel. Particular 
locations in the spectrum are labeled.  The first panel includes data
from across the whole line profile and a portion of the  continuum.
The next two panels show the behaviour across the P Cyg absorption, the
polarization changes with a relatively constant $\Theta$ of 20\degree\
and returns to the continuum polarization as the wavelength increases. 
Then, with increasing emission, the polarization changes with an
opposite sign but at the same angle until the central reversal in the
\ha\ line at --80 \km\ is reached at (0\%,--1.4\%).  From then (next
panel) the angle changes somewhat but the polarization reaches the same
values as for the P Cygni absorption at the red emission peak at +20
\km, to return to the continuum value as the emission wing reaches the
continuum. 
\label{qu}
}
\end{figure*}

Fig.~\ref{qu} shows the \ha\ spectrum in (Q,U) space.  For
clarity, several cuts along the velocity range (LSR) have been made. 
The first panel shows the (Q,U) vectors over the velocity range (--2500
\km,2500 \km) covering the \ha\ line and part of the continuum.  The
clustering of points at (1\%,--0.5\%) represents the continuum points,
while the polarization across \ha\ follows a loop in the
(Q,U) plot.  The subsequent panels break this loop up into segments.

The presence of the loop allows us to measure the position angle of
the scattering material with relative ease in (Q,U) space.  As the
contribution from circumstellar dust (and interstellar polarization)
amounts to a constant offset in the figure, the position angle can be
measured from the slope of the loop.  The overall angle $\Theta$ =
0.5$\times$atan($\Delta$U/$\Delta$Q) $\sim$ 20\degree\ can be
identified with the intrinsic polarization angle of the scattering
material.

\subsection{Interstellar polarization}

The main uncertainties in determining the intrinsic polarization of an
object are caused by the unknown value of the superimposed
interstellar polarization (ISP).  An additional problem in our case is
the unknown circumstellar dust polarization.  In the following we
discuss the ways in which it is possible to obtain a value of the ISP.

\subsubsection{Field stars and polarimetric variability}

Assuming that stars in the neighbourhood of the target star are not
intrinsically polarized, one can in principle argue that their
observed polarization is representative of the ISP suffered by the
target.  To this end, we have searched the polarization catalogue by
Matthewson \ea\ (1978).  Within a radius of 200\arcm\ of HD 87643, 59
objects were found.  However, their polarization angle and the
polarization percentage vary considerably.  The average polarization
and angle of the total sample is 1.34 $\pm$ 0.82\% and 119 $\pm$
32\degree.  We selected several subsamples of objects using either
\Av\ or the distance modulus as criteria, but the standard deviations
on the means did not improve significantly.  Consequently, the field
stars are best used as a consistency check rather than as the direct
means for determining the ISP.  Furthermore, this technique can not
provide a correction for the circumstellar dust contribution to the
polarization.

The polarimetric variability is an alternative method to separate the
intrinsic polarization from the ISP.  As the variable polarization is
due to the intrinsic polarization, one expects the polarization to move
along a straight line in (Q,U) space, thereby defining the intrinsic
polarization angle.  Broad-band polarimetry is provided by Barbier \&
Swings (1981), Gnedin \ea\ (1992) and Yudin \& Evans (1998). 
However,  the number and quality of the data-points is
not sufficient to improve upon the 20\degree\  found earlier.

\subsubsection{The `unpolarized' \ha\ method}

The field-star method and the polarimetric variability serve to
illustrate the problems commonly encountered in correcting for the ISP
superposed on polarimetric data.  Our last resort is to assume that
the \ha\ emission itself, at some wavelength within the profile is
intrinsically unpolarized.  Since this does not necessarily have to be
the case, there is clearly scope for this to be an
over-simplification.  However, this method also has the advantage that
a correction is made at once for both the circumstellar dust
contribution and the interstellar component.

Usually, the correction is made by measuring the polarization at the
line emission peak.  But as can be seen from Fig.~\ref{pol}, the line
peak is at a wavelength where the polarization is very sensitive to
wavelength, making it difficult to choose a well-constrained ISP
correction.  Fortunately, we know from the slope in the QU diagram
(Fig.~\ref{qu}) that the intrinsic position angle should be close to
20\degree.  We may also assume that the \ha\ emission is less polarized
at the systemic velocity, given the narrow nebular component of emission
associated with this object that is signalled by the [O {\sc i}] lines. 
We therefore chose to use an iterative procedure to arrive at the most
likely value of the ISP.  We derived the intrinsic polarization for
every point in  QU space close to the systemic velocity, assuming
that the observed polarization is due to interstellar and circumstellar
dust.  For all points close to the local minimum in the emission line,
we found resulting intrinsic polarization angles  close to
20\degree. The point at (Q=0\%, U = --1.2\%) resulted in an
intrinsic angle of 24 $\pm$ 1\degree.  This point corresponds to an
interstellar polarization of 1.2\% at an angle of 135\degree, which is
in fair agreement with the field stars (see above).  The wavelength is
40 \km\ (less than one resolution element) blueshifted from the systemic
velocity.  We will adopt this point as representative of the superposed
dust polarization. 

\begin{figure}
\includegraphics{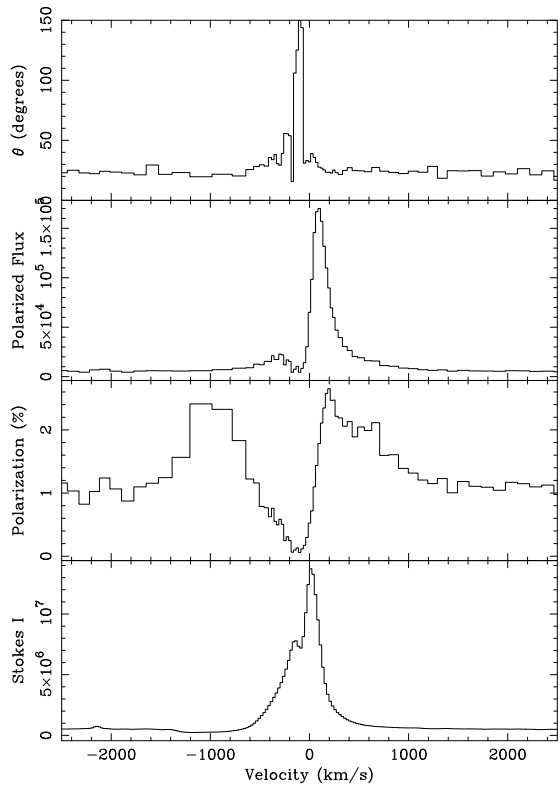}
\vspace{11cm}
\caption{
The intrinsic polarization spectrum around \ha\ obtained by correcting
the observed spectrum for a combined interstellar and circumstellar
polarization vector (Q,U) = (0,--1.2\%). 
\label{depol}
}
\end{figure}

The resulting intrinsic polarization spectrum is shown in
Figure~\ref{depol}.  The polarization runs much more smoothly along
the line profile, suggesting that the ISP has much to do with the
appearance of the observed spectrum.  If the ISP is measured in points
close to the chosen QU point, the resulting intrinsic polarization
spectrum is almost unchanged.  The polarization level across the
spectrum changes by less than few tenths of a percent, and the
rotation angle in the continuum changes by a few degrees. However,
the rotation angle through the \ha\ line centre changes considerably;
the peak angle changes from 150\degree\ for (0,--1.2\%) to 45\degree\
for (0,--1.4\%).

The P~Cygni absorption component still exhibits a higher percentage
polarization than the adjacent continuum, although there is no change in
polarization angle across this part of the line profile.  Despite the
problems in obtaining the intrinsic \ha\ polarization spectrum, it is
clear that the line-wings are more polarized than the continuum and that
the overall wavelength dependence does not simply mimic the intensity
profile.

\begin{figure}
\mbox{\epsfxsize=0.49\textwidth\epsfbox[100 80 580 480]{
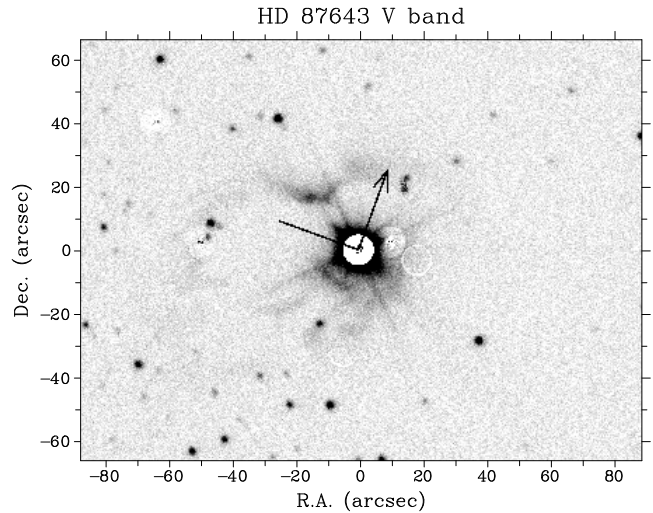
}}
\caption{
The nebula of HD 87643 in the {\it V} band. North is at the top, while
East is to the right.  The brightest field stars have been removed, and
the central star has been masked to enhance the faint structures. 
The arrow represents the intrinsic polarization  angle of 20\degree. 
\label{plaat}
}
\end{figure}

\subsection{The geometry of HD 87643 on larger scales}

The extended nebulosity of HD 87643, observed in the {\it V} band is
shown in Fig.~\ref{plaat} where the nebula is readily visible.  
Extended nebular emission was detected in both the {\it R} and {\it V}
images, and was more intense in the latter.  Similar nebular structures
around HD 87643 were not detected in the narrow-band H${\alpha}$
and [S\,{\sc ii}] images.  This confirms the reflection status of the
material as suggested by Surdej et al.  (1981). 

The geometry of the nebula is quite complex. Several components of the
nebula are visible.  There is an extension in the NNE direction, which
seems to have a counterpart, albeit much closer to the star, in the
SSW direction.  This may be evidence for multiple mass outbursts, but
deeper images are needed to appreciate such structures around the
star.  Perpendicular to this NNE-SSW component, is a larger extension
which is only present in the WNW direction.  The NNE-SSW component
seems to be aligned with the position angle of the intrinsic
polarization vector, with the western structure almost perpendicular
to it.  It may very well be that the circumstellar material around HD
87643 repeats the same geometry on both large scales (the reflection
nebula) and small spatial scales (seen in the \ha\ polarization).

\section{On the nature of and distance to HD 87643 }

As the distance to HD 87643 and, by implication, the star's evolutionary
status, is not yet a settled question, a  reassessment of its value
is appropriate.  We will show below that, based on the magnitude of the
interstellar extinction and kinematics, HD 87643 is more likely to be
an evolved object than a young star.

\subsection{Magnitude of interstellar extinction}

The extinction to HD 87643 can be determined in several ways: 

\begin{enumerate}

\item{For a spectral type of B2V, the total \ebv\ found by de Freitas
Pacheco \ea\ (1985) is 0.63 mag.  This value should not be greatly
dependent on the assumed spectral sub-type within the early-B
range. However, given the problems that inevitably arise in fitting
the spectral energy distributions of significantly reddened stars, 
this issue may benefit from a fresh examination}

\item{ The strength of the Na D absorption components (total EW 0.75
and 0.55 \ang\ for the D2 and D1 lines respectively) would indicate a
relatively large interstellar \ebv.  One should however be cautious in
interpreting the total EW of Na D absorption.  In a recent study of
the strength of the Na D1 absorption as a function of reddening by
Munari \& Zwitter (1997), the EW shows a linear dependence on
reddening at small EWs, while the dependence flattens for larger EW,
in a sense, this is a simple curve-of-growth effect, where saturated
lines need higher column densities than the linear part would imply.

Taking into account the multiple components in the Na~D lines we
arrive at a low \ebv\ of 0.2 $\pm$ 0.15. The caveat in this approach
may be that circumstellar emission components would artificially give
the impression of multiple components. However, taking the data as
they are, we arrive at a low value for the \ebv .

}

\item{The few Diffuse Interstellar Bands (DIBs) that are visible in
our spectrum suggest a larger \ebv.  The EW of the $\lambda$5780 DIB
is 0.3 \ang, which corresponds to \ebv\ = 0.51, based on the
calibration EW-\ebv\ provided by Jenniskens \& D\'esert (1994).  Other
DIBs in the spectrum result in the same values for \ebv:
EW($\lambda$5795+5797) = 0.18 \ang\ ( $\rightarrow$ \ebv\ = 0.72) and
\lam 6284 (combined 6281 and 6284; 1.05 \ang\ $\rightarrow$ \ebv\ =
0.57) ).  Since DIBs are weaker in the spectra of stars of which it is
known that much of the line-of-sight extinction originates in their
circumstellar shells, it is generally accepted that much of the
extinction traced by DIBs is interstellar (see the discussion by
Oudmaijer, Busfield \& Drew 1997, and references therein).  Assuming
that this applies to HD 87643 as well, the interstellar extinction
towards HD 87643 traced by the DIBs has a value \ebv\ $\sim$ 0.6, or
an \Av\ $\sim$ 1.9, for a ratio of total-to-selective extinction, \Rv
, of 3.1. This agrees well with the total \ebv\ as de Freitas Pacheco
\ea\ found based on the reddening of the spectral energy distribution,
but does not agree with the lower \ebv\ based on the Na D1 line.

}

\end{enumerate}

\begin{figure}
\mbox{\epsfxsize=0.48\textwidth\epsfbox[20 160 580 550]{
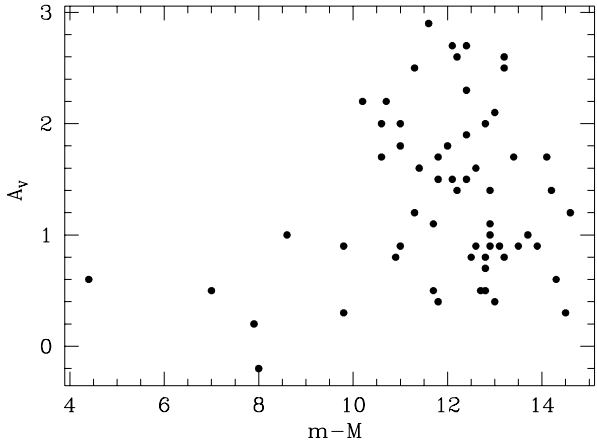
}}
\caption{
\Av\ as function of distance modulus for all stars in the Matthewson
\ea\ (1978) catalogue of polarized stars within 200\arcmin\ from HD~87643.
\label{mm}
}
\end{figure}

From the above, we find that the determination of the interstellar
extinction to HD 87643 is ambiguous, the DIBs would imply a \ebv\ of
order 0.6, while the Na D components may imply a lower \ebv.  This may
be compared with the interstellar extinction towards nearby stars.
The polarization catalogue compiled by Matthewson \ea\ (1978),
discussed earlier, also contains values for the extinction and
photometric distances of more than 7500 objects.  For the stars within
200\arcm\ from HD 87643, we find from Fig.~\ref{mm} that \av\ of order
1.9 suggests a distance modulus of order 10 -- 14 (1 -- 6 kpc), while
for lower values of \av\ the distance modulus can be anything between
0 and 15 magnitudes.

\subsection{Kinematic information}

Let us now consider the location of the object in the Galaxy.  The
object projects onto the sky close to the tangent edge of the Carina
spiral arm in the outer Galaxy.  At this particular longitude, the
distance to the near side of the Carina arm is estimated to be of
order 1.5 -- 3 kpc, while its far side extends to more than 10 kpc
from the Sun (Cohen \ea\ 1985, Grabelsky \ea\ 1987).  The CO and H
{\sc i} maps of Grabelsky \ea\ (1987) indicate that the intervening
local material, the near side and far side of the Carina arm are
relatively well separated in velocities.  The local material has LSR
velocities roughly between --9 to 7 \km, while the near side of the
Carina arm, overtaking the Sun, is moving at more negative velocities
(at least to $\approx$ --35 \km\ at the longitude of HD 87643), and
the far side, trailing the Sun, is located at more positive velocities
(to $\approx$ + 40 \km).

\begin{figure}
\mbox{\epsfxsize=0.48\textwidth\epsfbox[20 160 580 550]{
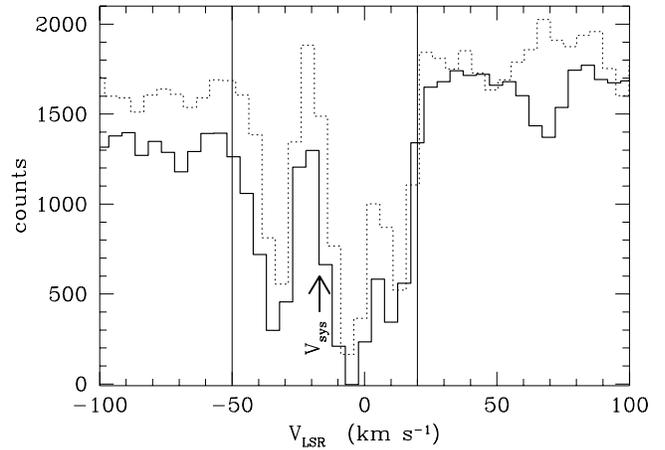
}}
\caption{
Na D1 \& D2 lines in the line of sight towards HD~87643
The systemic velocity of the star is also indicated.
\label{nad}
}
\end{figure}

Assuming that the Na D absorption is interstellar, we can use the
kinematics of the Na D absorption components to estimate the location
of the star in this sightline.  The Na D1 and D2 lines from the
echelle spectrum are shown on top of each other in Fig.~\ref{nad}.
Three distinct absorption components can be identified at --35 \km,
--9 \km\ and +10 \km\ in the LSR frame.  The vertical lines indicate
the entire velocity coverage, which ranges from --50 to +20 \km.  The
range of velocities spanned by the Na D absorption suggests that HD
87643 lies in the nearer half of the Carina arm.

If the interstellar extinction, \Av , is greater than 1, a broadly
consistent picture of HD 87643 at a distance of $\simgreat$ 2 kpc
emerges.  For an average {\it V} magnitude of 9, the intrinsic \mv\
lies in the range --2.9 to --6.9 for a distance range 1 -- 6 kpc.  The
bolometric correction for an early B-type star is of order --2.5
(Strai\v{z}ys \& Kuriliene 1981), so the bolometric magnitude of
HD~87643 would range between $\sim$ --5.4 to --9.4 (corresponding to
luminosities between 10~000 and 400~000 L$_{\odot}$ ).  The higher
luminosities are in the neighbourhood of those of B-type supergiants,
but the lower end can be occupied by both main sequence objects and
lower luminosity B[e] stars in the Magellanic Clouds (Gummersbach,
Zickgraf \& Wolf 1995). There is even a slight possibility
that HD 87643 is a low mass post-AGB star.

\subsection{HD 87643 as an evolved B[e] star}

The balance of probabilities on the question of the distance to HD
87643 is in favour of its being a (somewhat) evolved luminous object,
rather than a very young star (an opinion previously
expressed by e.g.  McGregor \ea\ 1988).  In this respect this star may
prove to be related to the so-called B[e] supergiants in the
Magellanic Clouds (see Zickgraf \ea\ 1985, 1986, 1996 and references
therein).  It is thus far not very clear whether these objects can be
linked to the Luminous Blue Variable phase of evolution, or how they
link to any other phase (e.g.  Schulte-Ladbeck 1998; Zickgraf 1998).

The B[e] stars exhibit hybrid spectra, with mostly very broad wind
absorption in the UV resonance lines, no photospheric absorption lines,
very broad hydrogen recombination emission lines, with absorption
components at small velocities, and a host of narrow
permitted and forbidden lines of singly ionized iron.  HD~87643 fits in
with this description.  The generally accepted model for these B[e]
objects was devised in 1985 by Zickgraf \ea\ (see also Zickgraf \ea\
1986).  It describes these objects as having a two-component wind.  Fast
outflows are thought to originate in a line-driven polar wind, while the
narrower lines are formed in a relatively slowly rotating, slowly
expanding circumstellar disk.  The presence of disk-like structures
around these objects was strengthened with the polarimetric observations
of Magalhaes (1992) which showed that the B[e] supergiants have strong
intrinsic polarizations. 

  HD 87643 stands apart from the Magellanic Cloud B[e] supergiants in a
number of respects.  Usually, fast wind component speeds up to $\sim$1000
\km\ are only visible in UV resonance lines, while the slow wind component 
is thought to be traced by the optical emission lines.  For example, the 
H$\alpha$ line profiles of the B[e] supergiants do not typically show P~Cygni 
absorption at velocities larger than $\sim$100 \km\ -- hence the absorption
is attributed to disk expansion alone.  The presence of both a broad
blueshifted and a narrow, nearly central absorption component in H$\alpha$
and H$\beta$ in HD~87643's spectrum is thus abnormal.  Somehow, in HD~87643
the fast-wind H{\sc i} column density is greatly in excess of typical
values.

  A further point of interest is that the forbidden lines are narrower
(FWHM of order 40 \km ) than the velocity typically inferred for the
`slow' winds (FWHM of 100 -- 150 \km\ for the Fe{\sc ii} lines). The
data of Zickgraf et al. (1986) for the LMC stars hint at different
outflow velocities as well, but not as clearly as for HD 87643.  This
indicates that low-density nebular emission constitutes a third
component which may, perhaps, trace the expanded remnant of a previous
mass loss phase of the star.  This third physical component might also
give rise to the far-infrared excess.

\subsection{Significance of the H$\alpha$ polarization}

    The complex behaviour of the linear polarization across the
H$\alpha$ line contains information, over and above
that available from direct spectroscopy, on the kinematics of the
immediate circumstellar medium of HD~87643.

Probably because of the limited use of high resolution
spectropolarimetry, only few instances have been reported in the
literature concerning enhanced \ha\ polarization, or changes along the
\ha\ line profile itself: the symbiotic system BI~Cru has an enhanced
\ha\ line polarization (Harries, 1996), Harries \& Howarth (1996)
found enhanced polarization in the \ha\ line centre of the O
supergiant $\zeta$ Puppis, while Schulte-Ladbeck \ea\ (1994) showed
that the line wings in the \ha\ line of the Luminous Blue Variable AG
Car have a different polarization signature than either the line
centre or the continuum.

A clue to the origin of the observed polarization characteristics of HD
87643 may be provided by model predictions of electron-scattering
geometries.  A start in developing such models has been made by Wood,
Brown \& Fox (1993), who calculated the polarization characteristics of
rotating and expanding disks around stars.  Although their schematic
model does not include radiation from the circumstellar material itself,
their results provide insight into the polarization expected from such
disks.

Wood \ea\ (1993) present the expected polarization due to a scattering
rotating disk - viewed at various inclination angles - and find that
this results in equally polarized blue and red emission, and, because
of the left-right anti-symmetry of the velocity field from the
observer's (and scatterer's) point of view, a flip in rotation angle
from one peak to the other.  In contrast to this, a purely expanding
disk results in enhanced polarization in the red emission peak -- a
response to the monotonic expansion of the atmosphere -- and a net
decrease in the blue wing.  There is no angle rotation in this case as
the velocity field seen by the observer is instead left-right
symmetric.  Neither of these cases describes, even qualitatively, the
polarization of H$\alpha$ in HD~87643.  However, in the case of a
hybrid velocity field in which rotation and expansion are both
present, Wood \ea's model results do come close to our observations.
The model shows enhanced polarization on both sides of the line
profile, with the red peak slightly more polarized than the blue peak,
while the angle rotation only takes place near line center.  The only
respect in which this pattern differs from that in HD~87643 is that
the blue and red peaks in the polarization percentage are almost
matched.

Encouraged by Wood et al's models, one may deduce, qualitatively from
the polarized flux spectrum (Fig.~\ref{depol}, in which the
blueshifted peak is much weaker than the redshifted) that the
scatterers `see' mainly expansion with only a modest component of
rotational motion mixed in.  For the timebeing this result on
H$\alpha$ points the way toward the correct dynamical model for B[e]
stars.  Eventually it will afford an important quantitative test.

\section{A radiation-driven disk wind model
for HD~87643 and other B[e] stars }

\begin{figure*}
\mbox{\epsfxsize=0.45\textwidth\epsfbox[50  -30 570 550]{
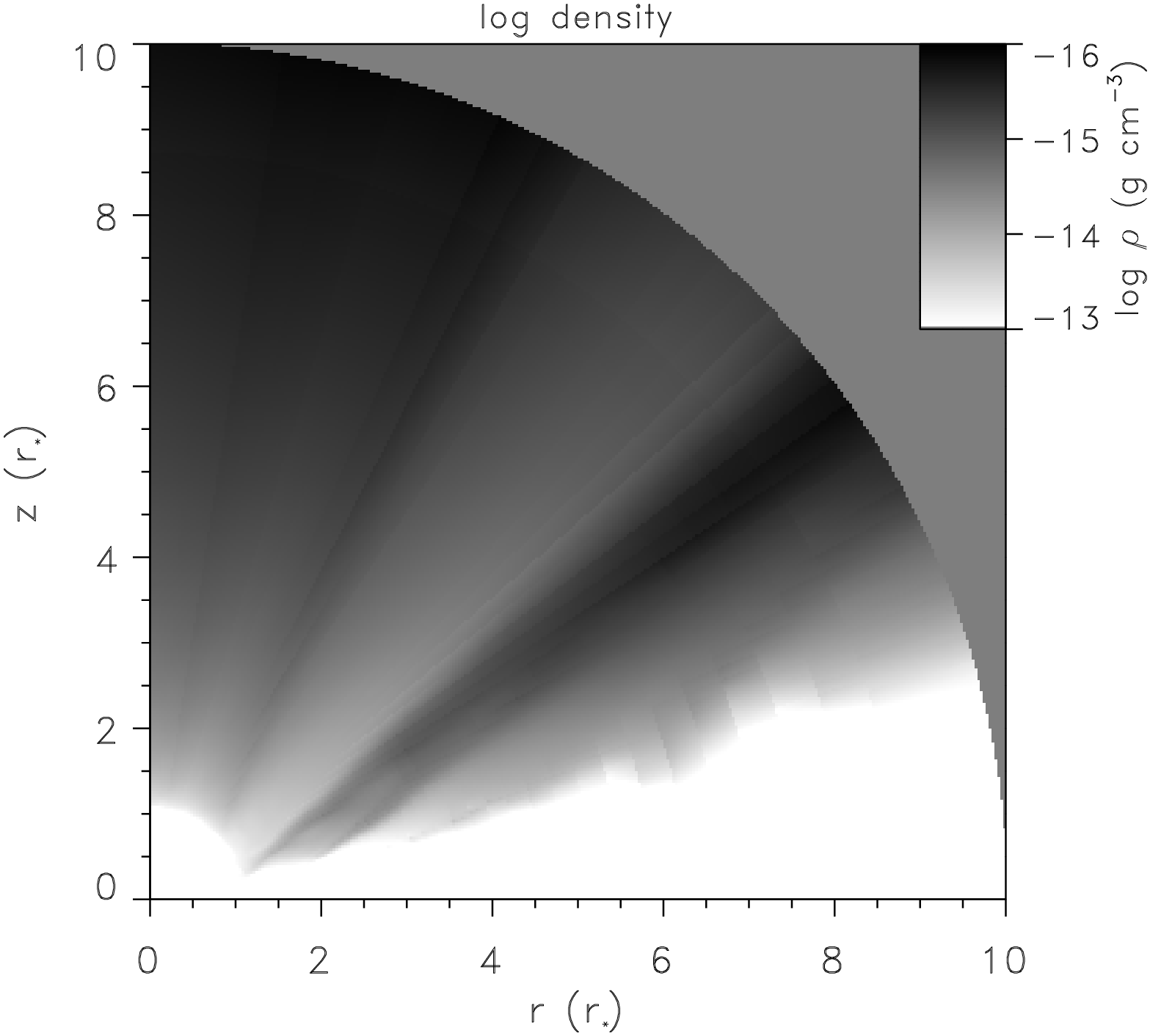
}}
\mbox{\epsfxsize=0.45\textwidth\epsfbox[-50 -30 470 550]{
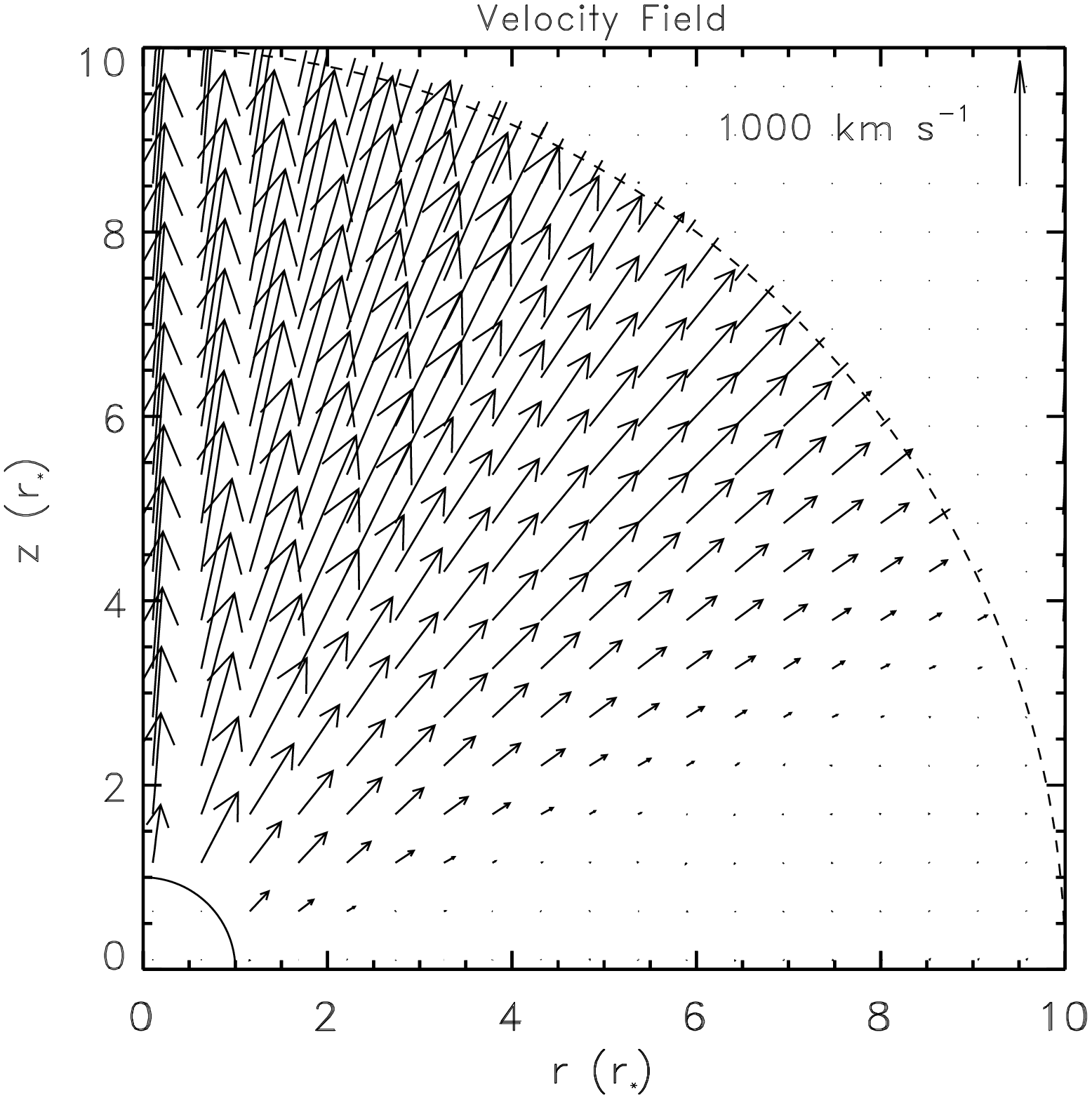
}}
\caption{
The grey-scale density map and velocity field for a model of a B[e]
supergiant with an accretion disk, described in the text (left and
right hand side panels). To make the density changes in the map more
visible, we set the disk density to be not higher than $10^{-13}~\rm
g~cm^{-3}$.  The rotational axis of the disk is along the left hand
side vertical frame, while the disk midplane is along the bottom
horizontal frame.  Note that the right hand side panel does not show
the rotational component of the velocity.
\label{den}
}
\end{figure*}

The two-component wind model of Zickgraf \ea\ (1985) explains the
observed characteristics of B[e] stars very well in a qualitative way,
without specifying the underlying dynamics.  The fast polar flow is
easily understood as a normal line-driven wind.  The difficult issue
that has stood in the way of a complete dynamical description is the
cause of the slow equatorial flow.  In recent years models have been
proposed in an attempt to deal with this, the focus being upon the
production of a slow wind direct from the surface of a rapidly
rotating star.  Here we set aside the basic difficulty of producing
the slow equatorial flow from the star along side the fast polar wind.
Instead we investigate the consequences of {\em assuming} there is a
Keplerian circumstellar disc that may also be a source of mass
loss. We then show that the model's properties offer considerable
promise of explaining what is observed in B[e] stars such as HD 87643.

Whatever the dynamical origin of a disk structure in B[e] stars and
similar objects, it is interesting to determine the system properties
when the presence of a disk is assumed a priori.  Proga, Stone \& Drew
(1998) have begun such a study.  Originally aimed at explaining the
properties of mass loss from cataclysmic variables, their hydrodynamic
model of radiation driven winds from accretion disks has much wider
application.  For example, Drew, Proga \& Stone (1998) have adapted
and extended the model to describe massive young stellar objects
(YSOs).

Here, we assume that an optically-thick disk exists around a B star
for reasons unknown, and that it shines by virtue of reprocessing the
stellar radiation falling upon it.  The star itself is non-rotating, a
convenient assumption rather than one that defines the model (see
later).  The disk is geometrically thin in the sense that the
reprocessed stellar radiation is emitted from the disk mid-plane.
Each point on the disk is assumed to emit isotropically.  The main
input parameters for the 2.5-dimensional line-driven wind model are
the stellar parameters: mass 25~\zm, radius 25~\rsun, and luminosity
$10^5$~\lm\ (a luminosity in the middle of the range discussed by
Gummersbach et al.  1995). Unless specifically mentioned, the model
parameters are as in Table 1 of Drew \ea\ (1998).

To calculate the structure of a wind from the star and the disk, we take into 
account stellar gravity, gas pressure effects, rotational and radiation
forces.  We hold the fluid temperature constant
at 15000~K in our isothermal equation of state.   
We use the CAK force multiplier (Castor, Abbott \& Klein 1975) 
to calculate the line-driving force.  In this approximation, a general form 
for this force is:
\begin{equation}
    F^{rad,l} = \int\!\!\!\!\! \int_{\Sigma} \left(\frac{\sigma_e d{\cal F}}{c}\right) 
         M(t) .  
\end{equation} 
The term in brackets is the electron scattering radiation force and
$M$, the force multiplier, is the increase in opacity due to spectral
lines.  The integration is over all visible radiating surfaces
($\Sigma$).  Our formalism allows the radiation from the central
accreting star to be included both as a direct contributor to the
radiation force and as an indirect component via disk irradiation and
re-emission.  Note that $d{\cal F}$ contains the total
frequency-integrated intensity emitted at any given location.  We
adopt the simple form for $M$ which still underpins much modelling of
OB star winds, i.e.  $M = kt^{-\alpha}$, where $t$ is proportional to
the local density divided by the local velocity gradient, and $k$ and
$\alpha$ are constants (taken to be 0.3 and 0.5 respectively).  These
calculations have been performed as described for massive YSO's
in Drew \ea\ (1998).  Some modifications are needed to
accommodate a lower gravity, higher luminosity B[e] star, these are
outlined below,

The boundary conditions are the same as in Drew \ea\ (1998; see also
Proga \ea\ 1998), except for the boundary density along the disk
mid-plane and stellar surface.  By holding the temperature of the gas
constant, the pressure scale height particularly in the outer parts of
the disk is artificially high.  In previous models the density, $\rho$,
along the disk mid-plane was independent of radius, $r$.  Here, the
combination of constant gas temperature with a markedly lower stellar
gravity gives rise to an over-thick disk atmosphere.  To be consistent
with the assumption of a geometrically thin disk in this circumstance,
it is necessary to modify the density in the disk mid-plane such that
it declines with increasing disk radius.  We choose $\rho \propto
r^{-\frac{21}{8}}$.  Such a scaling of the boundary density with
radius mimics the radial gas pressure gradient appropriate to a
steady-state disk (e.g.  Carr 1989).  For the disk boundary density at
one stellar radius, we adopt $10^{-9}~\rm g~cm^{-3}$, which is also
the density assumed along the stellar surface.

Because the stellar luminosity in our model is about 10\% of the
Eddington luminosity, we have to take steps to ensure that the
radiation pressure due to electron scattering does not, unphysically,
take effect inside the dense disk where in reality it should be
negligible in the presence of a nearly isotropic radiation field.  It
is not negligible only because of our simplification that the
radiation field streams freely from the disk and stellar boundaries.
We have found that this is most problematic in the inner disk where
the radial force term due to the stellar flux can be large if it is
not corrected for radiation transfer effects.  The device used to
remove this difficulty is to attenuate the stellar flux in the radial
electron-scattering radiation force integral by multiplying it by
$e^{-\tau_{es}^r}$, where $\tau_{es}^r$ is the optical depth in the
radial direction.  We approximate the optical depth by

\begin{equation}
\tau_{es}^r = \sigma_e \, \rho_{Disk} \,  \delta r_{Disk}
\end{equation} 

\noindent
where $\sigma_e$ is the mass scattering coefficient of free electrons,
$\rho_{Disk}$ is the density of the disk in hydrostatic equilibrium,
and $\delta r_{Disk}$ is the pressure scale length of the hydrostatic
disk. We calculate $\delta r_{Disk}$ using Proga \ea's equation (11)
defining it to be the density $e$-folding length at fixed polar angle,
$\theta$.

Finally, it should be emphasised that the disk in our model is best
thought of as simply a reservoir of material in Keplerian
orbits that is also optically-thick and therefore capable of intercepting
and re-radiating the stellar light falling on it.  In principle, the
radial density gradient across the disk surface can be chosen at will
if all orbits within the disk are viewed as non-interacting.  We have
chosen instead, partly for pragmatic reasons, to adopt a steady-state
configuration that would be appropriate to either accretion or excretion.

\subsection{Results of the model}

Figure~\ref{den} shows the grey-scale density map and velocity field
of our model.  Our hydrodynamic calculations show that the radiation
pressure due to lines produces a fast stellar wind in the polar region
and a slow wind near the equator.  
Thus, this wind model very
much resembles the two-component wind model proposed by Zickgraf \ea\
1985 (see also Zickgraf \ea\ 1989).  The main difference between
Zickgraf's conceptual model and our quantitative dynamical calculation
is that the latter incorporates a zone at intermediate latitudes
wherein the stellar outflow is compressed both because of the volume
taken up by the disk wind component, and because of the additional
component of vertically-directed line radiation pressure contributed
by the disk.  In this region the wind is slower and denser than at the
pole (see below).

Figure~\ref{den2} presents the run of the density, radial velocity,
mass flux density, and accumulated mass loss rate as a function of the
polar angle at the outer boundary, $r_0$, of our model (a distance of
10 stellar radii from the star center). The accumulated mass loss
rate is formally:

\begin{equation}
   d \dot{m}(\theta)  =  
       4\pi r_0^2 \int_{0}^{\theta} \rho v_r \sin\theta {\rm d} \theta
\end{equation}

The density contrast between the polar regions and $\theta \sim 70^o$,
where the radial expansion velocity is $\approx 100~\rm km~s^{-1}$, is
about 2 orders of magnitude.  The equatorial wind, originating from
the disk, is clearly very much denser and more slowly expanding than
the polar flow originating from the star.  The mass-loss rates in the
two flow components are comparable (see Fig.~\ref{den2}).  On the
other hand, the stellar wind compressed region at $\theta \sim 45^o$,
is denser by a factor of $\sim$ 5 and slower by a factor of $\simless$ 2
than in the polar regions.

\begin{figure}
\mbox{\epsfxsize=0.46\textwidth\epsfbox[70 310 550 700]{
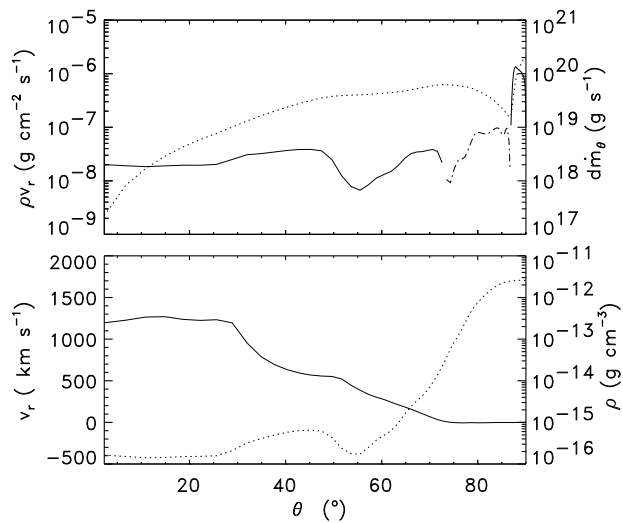
}}
\caption{
Quantities at the outer boundary in the model from Figure~\ref{den}.
The ordinate on the left handside of each panels is marked by the
solid line, while the oridinate on the right hand side is marked by
the dotted line.  The three components of the ouflow are easily
distinguished on this figure:1) the `normal' stellar wind is for
$0^o\simless \theta \simless 20^o$, 2) the disk compressed wind is for
$30^o\simless \theta \simless 50^o$, 3) the slow disk wind is
$50^o\simless \theta \simless 70^o$.  The zone for $\theta \simgreat
70^o$ is where the density profile only slightly deviates from the
hydrostatic profile and the matter is a subject of slow velocity
oscillations.  The gap in the mass flux density is caused by a
negative value of $v_r$ for $73^o\simless \theta \simless 87^o$. We
filled the gap by plotting $-\rho v_r$ with a dot-dashed line.
\label{den2}
}
\end{figure}

\subsection{Implications of the model}

The model calculations yield some interesting results which may begin
to explain some of the puzzles in the spectra of HD~87643 and B[e] stars
in general. 

In the hydrodynamic model described above, irradiation of the
optically-thick disk by the star is dynamically important.  Without
re-processing starlight, the disk on its own could not have promoted
outflow from its surface.  An effect of irradiation we have not had to
treat explictly in the hydrodynamics is the modification of the
emergent spectral energy distribution (SED).  Let us now anticipate
what this might be.  Illustrative calculations by Kenyon \& Hartmann
(1987) showed that the main change due to irradiation on the overall
SED is extra emission longward of the stellar Planck maximum.  The
excess fluxes at the {\it J} band found by Zickgraf \ea\ (1985) may be
evidence of this.  For B[e] stars, the disk's contribution to the
observed continuum can be significant also at optical wavelengths.
For example, when a completely flat circumstellar disk around an early
B star is viewed close to face on, it may contribute up to half of the
$B$ band light (see Figure 1 in Drew et al. 1998).  A disk of finite
opening angle would contribute more.  This veiling can go some of the
way to explain the observed lack of photospheric absorption lines 
in HD 87643 and other B[e] stars.  In addition, the externally heated
disk atmosphere is likely to be a source of heavy element line
emission (rather than line absorption) that could serve to cancel out
much of the stellar photospheric absorption.  For example, Hubeny
(1990) shows how strongly irradiated disk atmospheres are
characterised by inverted temperature profiles that  predispose
to net spectral line emission.

A further consequence of the irradiated disk scenario is that the disk 
can provide a much larger source of background continuum emission which
the wind has to shadow to produce net blueshifted absorption in a line
such as H$\alpha$.  This fact may explain why, in H$\alpha$, the 
high-velocity absorption is almost flat-bottomed, steep-sided and yet far 
from black.  To grasp this, recall an old result in stellar atmosphere theory 
that the absorption in an optically-thick purely resonantly scattering 
transition achieves a maximum depth of half the continuum level (see also 
Drew 1985).  This situation can develop because the line source function is 
then $WI_c$, where $I_c$ is the background continuum intensity, $W$ is the 
dilution factor and of course $W \geq 1/2$ within the atmosphere.  The same 
general idea may have some relevance to H$\alpha$ in HD~87643 in that the 
$n=2$ level of neutral hydrogen may be well enough populated that resonant 
scattering is significant, and that the line is forming against a larger than 
stellar continuum source -- with the consequence that the typical dilution 
factor in the line-forming region is not much below $1/2$.

The new model provides an improved overview as to what may be going on
in the line spectrum as a whole.  The strong spectropolarimetric
signature demands that our viewing angle must be reasonably high.  The
slowly expanding disk wind component would need to be the origin of
the central part of the H$\alpha$ line profile, i.e. the double-peaked
emission and the asymmetric dip in between.  The somewhat broadened
Fe~{\sc ii} lines (e.g. in Figure 4) will come from near the base of
the disk wind (in keeping with Zickgraf's original concept).  The
absence of a strong rotational signature in any of the observed line
profiles can be accounted for by two cooperating factors -- optical
depth effects and angular momentum conservation.  The majority of
lines may be optically-thick, implying that the observed emission
samples the outflow (where there is little residual rotation) well
away from the star and the disk mid-plane.  However, we do know from
the H$\alpha$ spectropolarimetry that some rotational motion is
detectable.  This is very important and very encouraging.

The high velocity blueshifted H$\alpha$ absorption would most likely
arise from the compressed stellar wind component.  An interesting point
about this feature, particularly apparent in the later April observation
(Fig.~\ref{ha}), is that it cuts on rather sharply at around $-800$
\km .  This suggests that the density of the absorbing gas is not  a
monotonic function of the flow velocity.  Our
hydrodynamic model shows that such irregularities of the density and
velocity are present.  As mentioned above, the compressed stellar wind
component has densities $\sim 5$ times higher and expansion velocities
about half those characterising the radial wind over the pole.  At
the interface between the compressed stellar wind and the slow disk
wind, there is at first a decrease (at fixed radius) in density with
increasing polar angle, and then there is a sharp rise again as the disk
atmosphere is approached (Fig.~\ref{den2}).  

Over this same angular range, the expansion velocity at fixed radius
steadily increases. There is thus an absorbing column of higher
density. higher velocity material at $\theta \sim 45^o$ along side a
column of lower density, lower velocity gas at $\theta \sim 55^o$.
Spectral line synthesis is required to investigate whether such a
configuration can match the observed \ha\ profiles in HD 87643.

\subsection{Relation to previous models}

    The fact of the strong contrast between the polar and equatorial
flows implied by observations of B[e] stars obviously demands an
underlying geometry that is axial rather than spherical.  The
introduction of rapid rotation of the star is one way of achieving this.
For example, Ignace,  Cassinelli \& Bjorkman (1996) showed that the 
`wind-compressed disk' model (Bjorkman \& Cassinelli, 1993) produced
for classical Be stars that are known to be rapid rotators, could also 
apply to B supergiants in the event that they too undergo significant 
rotation.  In this framework, the disk in Zickgraf's concept is seen as 
a wind compressed zone.  Another mechanism that may enable a rapidly
rotating star to produce a dense equatorial flow alongside a fast polar
wind is the `bistability' of radiation-driven winds discussed by
Lamers \& Pauldrach (1991).  In this case, the pole/equator contrast 
arises from latitudinal differences in wind ionization and optical depth.
Owocki, Gayley \& Cranmer (1998) have reviewed
the impact on both models of gravity darkening and non-radial line
forces. They conclude that non-radial line forces may prevent the 
formation of the wind-compressed disk in rapidly rotating hot-star winds.
They also provide an expression for the equator/pole density contrast 
due to bistability operating in the presence of gravity darkening.  
Using Owocki \ea 's (1998) numbers with $\theta > 70^{\rm o}$ and a
stellar rotation 80\% of critical in this expression, we obtain a density 
contrast of around an order of magnitude at best.

    Whilst it is clear that there is more work to be done to confirm
the outcome of Owocki \ea 's (1998) analysis, it is certain that
stellar rotation on its own cannot {\em easily} account for the high
equator/pole density contrast thought to characterize B[e] stars.  
However, because our model assumes a significant Keplerian
circumstellar disk is already in place, a density contrast of 2
magnitudes or more naturally follows.  If the star in our model were a
rapid rotator, the important change with respect to the present
non-rotating case would be weaker irradiation of the optically-thick
disk.  In turn this would imply a reduction in the mass loss rate,
together with a lower opening angle for the disk wind.  Combined,
these effects should result in only a small change to the equator/pole
density contrast.  It is in this sense that stellar rotation is a
second-order effect in our calculations.

   However, it is a distinct possibility that an optically-thick
circumstellar disk only arises where the star is a rapid rotator --
the situation envisaged by the wind-compressed disk and bistability
models.  If these models do not survive continuing scrutiny, it may
prove to be the case that another dynamical process enables B[e] stars
to generate optically-thick Keplerian disks around themselves.  Given
the uncertainty over the evolutionary status of B[e] stars, and the
evidence that some may be quite low luminosity objects (Gummersbach et
al 1995, see also Drew \ea\ 1997 on MWC~297), it is also a possibility
that some B[e] stars are still in possession of disks created in an
earlier evolutionary phase.  For example, some may be extreme post-AGB
stars that in time will produce bipolar nebulae.  Some may be B stars
that have not yet dissipated the accretion disks from which they
formed.  Indeed this could be a solution to the longstanding puzzle,
discussed by Herbig (1994), of the phenomenological similarity between
B[e] stars and young BN-type objects -- perhaps they look the same
because they are the same, and disk dissipation timescales are long
enough to allow this.  The main point of our disk-wind model is that
it explores the consequences of an optically-thick circumstellar disk
around a B star without first solving the harder problem of its
origin.

\section{Concluding remarks}

In this paper we have presented an observational study of the B[e] star
HD 87643.  The observations include high dispersion spectroscopy, medium
dispersion spectropolarimetry around \ha\ and imaging.  
The main conclusions drawn from them are as follows: 
 
\begin{enumerate}

\item{From the spectroscopic observations we have found that at least
three different line forming regions contribute to the spectrum; a fast
outflow giving rise to extremely high velocity components up to 1800
\km\ in \ha\ and neutral helium, a slower outflow visible in the
relatively low velocity central reversal absorption in the hydrogen
lines and in broad (FWHM $\sim$ 100 -- 150 \km) \al{Fe}{ii} lines, while a
third, narrow, component (FWHM $\sim$ 40 \km) is revealed by narrow
forbidden lines, neutral \al{Fe}{i} lines and some weaker \al{Fe}{ii}
lines. 
}

\item{ A disk-like structure surrounding the star is revealed by the
medium resolution spectropolarimetry.  The startling linear polarization
changes across the \ha\ emission line immediately imply that there is
scattering within the ionized line-forming region and that this volume
is certainly not spherical.  A comparison with published model
calculations shows that a rotating, expanding disk qualitatively
reproduces the observed polarization, and illustrates the need for more
realistic model calculations. 
}

\item{ Broad-band imaging of the large reflection nebula around HD
87643, suggests that the geometry of the circumstellar material at large
scales are comparable to those on the small scales traced in the
polarimetry; the intrinsic polarization angle aligns well with the two
most conspicuous nebular components. 
}

\item{Interstellar extinction and kinematic evidence favours a
distance to the object of order of several kiloparsecs.  This implies
that it is unlikely that HD 87643 is a young main sequence object.
Nevertheless, these data are not conclusive in deciding whether HD
87643 is a luminous B-type supergiant or a lower luminosity type B[e]
star such as those discussed by Gummersbach \ea\ (1995).  Indeed a
feature peculiar to HD~87643 is that the maximum outflow velocities
seen in absorption in the H$\alpha$ line ($\sim$ 1800 \km ) are
significantly higher than those seen in other B[e] stars luminous
enough to be classified as supergiants.  Since this absorption can
only arise in the stellar wind and so may be sensitive to the
stellar photospheric escape velocity, a more compact star with a
luminosity significantly below that of a supergiant is a possibility. }

\end{enumerate}

These results on HD~87643 fit into the general conceptual framework due to 
Zickgraf \ea\ (1985,1986) in which B[e] stars are described as the source of 
two winds -- a fast wind in the polar direction and a slower equatorial 
disk-like wind.  Here we have proposed a model that 
provides a possible physical basis for Zickgraf \ea 's two-wind concept.  
Our model consists of a star encircled by an {\em optically-thick} disk 
wherein radiation pressure is sufficient to power mass 
loss from both the star and disk.  Hydrodynamical calculations yield a fast 
wind emerging from the star's polar regions and a much slower wind flowing 
away from the disk.  A feature of the calculations not anticipated in earlier 
qualitative B[e] star models is that the combination of the irradiated disk 
and its outflow modify the stellar wind at intermediate latitudes, rendering 
it denser and slower than over the stellar pole.  The new picture these
calculations present offers considerable promise of explaining our observations
of HD 87643.  An important next step will be to synthesise spectral line
profiles from the model in order to make a direct quantitative comparison
with observations.

\paragraph*{\it Acknowledgments}

Anna Gatti is thanked for her help in obtaining the MSSSO spectra. 
We thank the staff at the Mount Stromlo and Siding Spring Observatories
(MSSSO) and the Anglo-Australian Telescope for their expert advice and
support.  The allocation of time on the Anglo-Australian Telescope was
awarded by PATT, the United Kingdom allocation panel.  Computations were
performed at the Pittsburgh Supercomputing Center.  RDO and DP are
funded by the Particle Physics and Astronomy Research Council of the
United Kingdom.  DdW is supported in part by Spanish grant DGICYT
PB94-0165. This research has made use of the Simbad database, operated at
CDS, Strasbourg, France.

\end{document}